\documentstyle[aasms4,psfig]{article}
\def\gs{\mathrel{\raise0.35ex\hbox{$\scriptstyle >$}\kern-0.6em 
\lower0.40ex\hbox{{$\scriptstyle \sim$}}}}
\def\ls{\mathrel{\raise0.35ex\hbox{$\scriptstyle <$}\kern-0.6em 
\lower0.40ex\hbox{{$\scriptstyle \sim$}}}}

\def\arcsper{\ifmmode \rlap.{''}\else $\rlap{.}''$\fi}
\def\arcmper{\ifmmode \rlap.{'}\else $\rlap{.}'$\fi}

\def\RR{\hbox{$R_{702}$}}
\def\II{\hbox{$I_{814}$}}
\def\VV{\hbox{$V_{555}$}}
\received{---}
\revised{---}
\accepted{---}

\begin{document}

\title{A COMPARISON OF DIRECT AND INDIRECT\\
MASS ESTIMATES FOR DISTANT CLUSTERS OF GALAXIES\footnote{
Based on observations 
obtained with the NASA/ESA Hubble Space Telescope
which is operated by STSCI for the Association of Universities
for Research in Astronomy, Inc., under NASA contract NAS5-26555.}}
\author{
Ian Smail\altaffilmark{1}\footnote{Current address: Department of Physics, 
University of Durham, South Rd, Durham DH1 3LE, UK},
Richard S.\ Ellis\altaffilmark{2},
Alan Dressler\altaffilmark{1},
Warrick J.\ Couch\altaffilmark{3},\\
Augustus Oemler Jr.\altaffilmark{4}\footnote{Current address: 
The Observatories of the Carnegie Institution of Washington, 
813 Santa Barbara St., Pasadena, CA 91101-1292},
Ray M.\ Sharples\altaffilmark{5} \&
Harvey Butcher\altaffilmark{6},
}

\bigskip
\affil{ 1) The Observatories of the Carnegie Institution of Washington,\\
813 Santa Barbara St., Pasadena, CA 91101-1292}
\affil{ 2) Institute of Astronomy, Madingley Rd, Cambridge CB3 OHA, UK}
\affil{ 3) School of Physics, University of New South Wales, Sydney 2052, Australia}
\affil{ 4) Astronomy Department, Yale University, PO Box 208101, New Haven CT 06520-8101}
\affil{ 5) Department of Physics, University of Durham, South Rd, Durham DH1 3LE, UK}
\affil{ 6) NFRA, PO Box 2, NL-7990, AA Dwingeloo, The Netherlands}

\begin{abstract}
We present  weak lensing results for 12 distant clusters determined
from images obtained with the refurbished {\it Hubble Space
Telescope}.  We detect the signature of gravitational lensing in 11 of
the 12 clusters; the clusters span nearly an order of magnitude in
lensing strength.  The sample thus provides an excellent database for
correlating direct mass estimates based on lensing with indirect ones
which  rely on baryonic tracers. We examine the correlation between the
cluster X-ray luminosities and the mean gravitational shear strengths
and develop a model which allows us to predict the relationship
expected from the properties of local clusters.  After allowing for
various observational effects, we find that the predicted correlation
is a reasonable match to the available data, indicating that there has
been little evolution in the X-ray luminosity--central mass
relationship between $z\sim 0.4$ and now. We discuss the implications
of this result in the context of the evolution of the X-ray luminosity
function found by earlier workers. The comparison between shear
amplitudes and velocity dispersions, estimated from a modest sample of
members ($\sim 30$), reveals a discrepancy in the sense that these
velocity dispersions are typically over-estimated by factors of $\sim
50$\%.  This supports earlier suggestions that high dispersions
measured for distant clusters may be seriously affected by both
unidentified substructure and outliers.  Combining our shear-based mass
estimates with morphologically-based luminosity estimates, we determine
mass/light ratios of $M/L_V^{all}$=180$^{+210}_{-110}h$ (M/L)$_{\odot}$
for the entire population and 620$^{+250}_{-240}h$ for spheroidal
population where the evolutionary effects can be best treated. We argue
that this provides an upper bound to the present-day cluster mass/light
ratio corresponding to $\Omega \sim $0.4.  Our results demonstrate the
important role weak gravitational lensing can play in the study of the
evolution of distant clusters, as the most direct and least biased
probe of their growth.
\end{abstract}


\section{Introduction}

Rich clusters of galaxies can be identified to high redshift and can thus be 
used as tracers of the evolution of structure. Moreover, as clusters 
represent the extreme tail of the fluctuation spectrum they provide a 
particularly sensitive probe of the form of the primordial power spectrum.  

The most widely used techniques for identifying distant clusters are
based on searches either for an overdensity in the projected galaxy
distribution (\cite{gho89}; \cite{wjc91}; \cite{post95}) or the
detection of X-ray emission from hot gas bound to the cluster potential
(\cite{ace90}; \cite{jp94}; \cite{rgb94}; \cite{fjc95}). However, the
relatively high abundance of distant optically-selected clusters to
$z\simeq$0.5 appears to contradict the X-ray survey results which
suggest a marked decline in the volume density of luminous X-ray
systems beyond $z\simeq$0.2. This conflict has led to the suggestion
(\cite{nk91}) that the evolution of the X-ray luminosities of distant
clusters need not simply reflect changes in the underlying potential
wells. By including a minimum entropy in the gas prior to it entering
the cluster, Kaiser was able to limit the compression the gas undergoes
during the cluster formation, significantly altering the X-ray
evolution.  If confirmed, this would seriously complicate the use of
X-ray observations for studying the evolution of the mass function of
distant clusters. It would also raise interesting issues for the growth
of structure in the universe and the thermal evolution of the hot X-ray
gas in clusters (\cite{nk91}; \cite{fjc95}; \cite{rgb96}).

The optical richness of a cluster is clearly more prone to projection
effects than the cluster's X-ray luminosity.  To try to reduce this
problem, a number of groups are undertaking spectroscopic surveys of
distant clusters to provide membership information, in addition to
estimating the cluster velocity dispersions.  However, rather than
using indirect baryonic tracers such as cluster galaxies or hot X-ray
gas, it would clearly be preferable to consider more direct measures of
the cluster mass.  The most direct measure comes from the analysis of
the weak gravitational lensing signal detected using the shapes of
field galaxies seen through the cluster.  The lensing method can
provide an absolute measure of the cluster mass and can be readily
applied to intermediate and distant rich clusters. At the same
time it is insensitive to the thermodynamical state of the baryonic
components in the cluster, a particularly important advantage
when considering clusters at high redshift which may not be fully
virialised.

We report in this paper on a study of the lensing properties of 12
distant clusters which have been imaged using  the {\it Hubble Space
Telescope} (HST).  The clusters span a wide range in redshift
($z=0.18$--0.58), X-ray luminosity and optical richness, and thus
comprise an ideal sample for investigating the relative evolution of
the mass, X-ray and galaxian properties of distant clusters. The key to
our study is the derivation of mass estimates from the cluster's
lensing effect on uniformly-selected  samples of background faint
field galaxies. By measuring the distortion of this population we obtain a
robust measure of the mass in the central regions of the foreground
clusters.  Using simple, empirical models of the clusters, we  then
compare the predicted X-ray luminosities and velocity dispersions,
calculated from our mass estimates and local scaling relations, with
the observed quantities to distinguish between models for the evolution
of the cluster properties.  We then proceed to use our lensing mass
estimates, along with morphologically classified samples of cluster
galaxies, to estimate the mass to light ratios of our clusters.

A plan of the paper follows. The cluster dataset and its reduction is
summarized in \S 2 and the lensing analysis and modelling described in 
\S 3.  We discuss the results and compare the various mass estimates 
in \S 4, where we also give the mass to light ratios (M/L) estimated from
our lensing analysis before summarizing our conclusions in \S 5.

\section{Data}

The bulk of the HST data used for this study were obtained as part of a
Cycle-4 WFPC-2 program to investigate the morphological evolution of
cluster galaxies. This sample comprises ten distant clusters which have
been the subject of an intensive ground-based campaign by ourselves and
other groups over the past decade. Data for two other clusters (Cl0440$-$02
and Cl0024+16) have been retrieved from the HST archive. The full dataset
does not constitute a ``complete'' sample in any respect, but is
ideally suited for our purpose since they span a wide range of optical
richness and X-ray luminosity and have been well-studied hitherto. As
we show below, the wide range in properties is essential to adequately
test for differential evolution between the various estimators of the
cluster mass. The important issue for our analysis is the fact that
these HST images are sufficiently deep to provide reliable information
on the weak lensing of background galaxies viewed through the clusters.

The cluster sample, together with relevant observational details, is
given in Table~1.  The filters used for the observations discussed here
are F555W (\VV), F702W (\RR) and F814W (\II).  The individual exposures
were generally grouped in sets of 4 single-orbit exposures each offset
by 2.0 arcsec to allow for hot pixel rejection. After standard pipeline
reduction, the images were aligned using integer pixel shifts and
combined into final frames using the IRAF/STSDAS task CRREJ.  We retain
the WFPC-2 color system and hence use the zero points from Holtzman et
al.\ (1995). The final images cover the central 0.4--0.8 h$^{-1}$
Mpc\footnote{We use $q_o=0.5$ and $h = H_o / 100$ km sec$^{-1}$
Mpc$^{-1}$.  This geometry means that 1 arcsec is equivalent to 1.88
h$^{-1}$ kpc for our lowest redshift cluster and  3.76 h$^{-1}$ kpc in
the most distant.} of the clusters (Fig.~1) to a 5$\sigma$ limiting
depth of $I_{814} \simeq 26.0$ or $R_{702} \simeq 26.5$--27.0.  In the
following analysis we use the reddest band available for a particular
cluster to identify objects and measure their shapes. The \VV\ images
are used primarily for determining colors, which aid in identifying
cluster members.

To catalog faint objects in these frames and measure their shapes we
used the Sextractor image analysis package (\cite{eb95}).  We adopt a
detection isophote equivalent to $\sim 1.3 \sigma$ above the sky, where
$\sigma$ is the standard deviation the sky noise, viz.\ $\mu_{814}=
25.0$ mag arcsec$^{-2}$ or $\mu_{702} = 25.7$ mag arcsec$^{-2}$
($\mu_{702} = 25.0$ mag arcsec$^{-2}$ for A2218, AC103 and AC118), and
a minimum area after convolution with a 0.3 arcsec diameter top-hat
filter of 0.12 arcsec$^{2}$.  Analysis of our exposures provides
catalogs of $\sim 800$ objects for each cluster across the 3 WFC chips.
We discard the smaller, lower sensitivity, PC fields as well as a
narrow border around each WFC frame in the following analysis.

To compare the lensing strengths of the various clusters we must
construct well-defined samples of background galaxies for which image
parameters can be measured with adequate signal-to-noise. For
simplicity in modelling we have adopted uniform magnitude limits across
the sample. The faint magnitude limit is determined by the depth at
which reliable images shapes can be measured in our shortest exposures.
This is $R_{702} = 26.5$, as set by the A2218, AC103 and AC118
exposures.  The bright limit is set by our desire to reduce cluster
galaxy contamination in the field samples for the most distant
clusters. Using colors, we can determine the point at which the
color-magnitude relation for the cluster ellipticals blends into the
background population. Fainter than this point the field population
dominates (see below). This corresponds to a bright limit of $I_{814} =
24.0$.  When converting between the \RR\ and \II\ limited samples, we
have assumed a typical color for the faint field population at these
depths of $(R_{702}\! -\!  I_{814}) \sim 0.5$ (\cite{irs95b}). 

Applying these limits yields a typical surface density of $\sim 95$
field galaxies per arcmin$^2$, in good agreement with that measured in
genuine `blank' fields ($\sim 95\pm10$ arcmin$^{-2}$) after correcting
for differences in the photometric systems (\cite{irs95b}). We thus
estimate that any residual contamination in our catalogs from faint
cluster members must be less than $\sim 5$--10\%.   The final sample
size in a typical cluster, after applying both the magnitude and the
area cuts (see below), is $\sim 180$ galaxies.

Additional cluster data is also given in Table~1. This includes
rest-frame velocity dispersions from Couch \& Sharples (1987), Dressler
\& Gunn (1992) and Le~Borgne et al.\ (1992).  These are estimated using
standard procedures (\cite{dan80}) using those authors' own redshift
catalogues without further corrections. We also give X-ray luminosities
in the 0.3--3.5 keV band,  taken from the literature (\cite{fjc};
\cite{h82}; \cite{ws}), when available.  For four of the clusters
(AC118, Cl0939+47, Cl0412$-$65 and Cl0054$-$27) no published data were
available and for these we measured luminosities from archival ROSAT
PSPC images of the clusters.  The cluster flux is measured inside an
optimal aperture, determined from the cluster's detection significance,
and then corrected to a total flux using a $\beta$-model with
$\beta=2/3$ and r$_c = $125 h$^{-1}$ kpc.  When required, we converted
the X-ray luminosities to the 0.3-3.5 keV band assuming a mean cluster
temperature of 7 keV; this is typically a correction of $\simeq$20\%.
We adopt a mean cluster temperature of 7 keV when calculating these
corrections.  No attempt has been made to remove emission arising from
either a central cooling flow or any faint embedded point sources.

\section{Analysis}

\subsection{Weak Lensing Measurements}

We now discuss our methods for measuring the cluster shear from the catalogs 
of faint objects described above. 

Two-dimensional shear maps have been constructed from the complex
shear:  $g = \epsilon e^{2i\phi}$, where $\phi$ is the position angle
of the major axis of a faint image and $\epsilon = (a-b)/(a+b)$ is a
shape estimator for an image of major and minor axes $a$ and $b$
(\cite{jpk95}). This quantity is calculated for every image in the
catalog and the shear estimates are binned into independent $20 \times
20$ arcsec cells and plotted as a vector field over the frame. A number
of our clusters show strongly coherent shear fields (Fig.~1) which
provide a very powerful and direct view of the cluster potential well
including its center, ellipticity and orientation.  In other cases the
coherence of the shear field is more difficult to discern and we must
assume symmetry, and in some cases a center for the shear field, before
we can proceed with our analysis. In these cases, we assume, following
clear trends observed with our stronger lenses, that the shear geometry
is roughly circular and centered on the brightest cluster member. 
 
The total shear strength of a cluster, $<\! g_1\!  >$, was calculated
using the average tangential shear ($g_1 = - \epsilon \cos (2 \theta)$,
where $\theta$ is the angle between the major axis and the vector
joining it to the lens center) for all the images contained within an
annulus $r=60$--$200 $h$^{-1}$ kpc around the lens center. The shear is
determined in annular sections between the two radii and corrections
applied for any fraction of the annulus falling outside the WFC
boundary. Errors are determined by bootstrap resampling of the data and
are thus likely to be liberal estimates. The outer annulus represents
the largest metric radius common to all the clusters which does not
involve a large correction for areas off the frame. The inner radius
has been chosen so as to reduce any underestimation of the shear in the
central regions, due to either our adopted shear estimator or the
difficulties in detecting faint lensed features against the halos of
the luminous central cluster galaxies.

To determine the contribution to the observed shear from systematic
effects in the HST optics, detectors, or the reduction method, we have
also applied our analysis to a similarly deep image of a blank field.
This archival data consists of deep \VV/\II\ WFPC-2 images of a blank
field from the ``Groth Strip''. The data was processed and object
catalogs generated in a similar manner as before.  The mean tangential
distortion using our standard annulus and magnitude limits was found to
be $<\! g_1\!  > = -0.008\pm 0.070$. The quoted error represents the
field-to-field scatter, which is similar to the uncertainty determined
by boot-strap resampling.

One final step is required to place all our shear measurements on a
comparative scale. Notwithstanding the uniform magnitude limits, the
different redshifts of the various clusters means that each experiences
a different proportion of foreground field contamination. Furthermore,
because lensing depends on the relative distances of the lens and
background source, even identical clusters at different redshifts will
produce different shear effects on the same background population. We
have therefore corrected all the observed shear signals to that
appropriate for a putative cluster placed at the mean redshift of our
sample, $<\! z\! > \sim 0.4$, lensing a population of galaxies at $z\gg
0.4$. 

To determine this correction we need to make the following assumption:
that the  background population viewed through all the clusters can be
well described by a single redshift distribution, $N(z)$.  Following
the inversion analysis of the cluster A2218 by Kneib et al.\ (1995), we
assume the no evolution form for our magnitude-limited sample, ignoring
possible differences between the \RR\ and \II\ selection.  This $N(z)$
has a median $z=0.83$ with only $\sim 10$\% of the field population at
$z<0.4$. As the correction is differential we do not expect residual
uncertainties arising from the assumed $N(z)$ to be significant.  

\subsection{Observational Factors}

Here we calculate the effect of those observational factors which may
degrade the amplitude of the observed shear signal. Such effects
include possible inefficiencies in the image analysis algorithm, image
crowding, low signal-to-noise in the image shapes and residual cluster
contamination. Earlier workers have estimated the correction from
observed to true shear to be around $\sim 2$--3 for ground-based
observations (\cite{nk95}; \cite{gw95}), where the majority of the
degradation arises from atmospheric seeing. Working with HST we can
thus expect a smaller correction.   

To estimate our observational efficiency we have used two approaches.
Firstly, we have taken HST images of deep blank fields and sheared them
by known amounts, added in sky noise to simulate the depth of our cluster
exposures and then re-measured the shear signal from these images.
Secondly, we have created artificial images matched as closely as
possible to the characteristics of our observations (e.g.\ magnitude
and scale size distributions of the field population, sky noise, etc.)
and introduced a known amount of coherent shear. By applying our
analysis technique we can then determine the efficiency of our shear
measurement.  ~From these experiments we determine a degradation of
$\sim 0.8^{+0.1}_{-0.2}$. 

Comparing the surface densities of faint galaxies in our fields to counts 
from blank fields, we estimate that the maximum contribution from faint 
cluster members will be less than $< 10$\%.  Assuming that these faint 
images are randomly orientated, they will degrade the observed shear signal 
on average by $\sim 5$\%.  Combining this with the estimated analysis 
efficiency we obtain a total efficiency of $\sim 0.75\pm0.20$.  This signal
degradation can now be included in our model predictions (see below)
when comparing them to the observations.

\subsection{Theoretical Predictions}

Before comparing the cluster masses derived from our shear measurements
with the  X-ray luminosities (L$_X$) and cluster velocity dispersions
($\sigma$), we need to predict the expected behaviour of the
L$_X$--shear and $\sigma$--shear correlations using simple models for
the expected evolution of clusters.  Although this is clearly a complex
issue, our aim in this paper is to adopt a simpler approach, based on a
number of elementary assumptions, seeing whether these are capable of
reproducing the correlations observed in our sample.

We first examine the expected X-ray luminosity for a cluster lens as a
function of its shear amplitude.  To estimate an upper limit on the
predicted shear for a given cluster mass, we consider the hardest
cluster potential likely:  nearly-singular isothermal spheres. Such
potentials are supported by the mass profiles and core radii that have
been determined from detailed modelling of highly-constrained cluster
lenses (\cite{jpk95}).  In addition, if we fit the radial shear
profiles in our clusters with the form, $<\! g_1\! > \propto r^\gamma$,
we obtain $\gamma = -1.0\pm0.4$, further supporting our assumption of
an isothermal mass profile. By including an inner annular cutoff our
results are relatively insensitive to the mass profile of the innermost
regions where strong cluster-to-cluster variations may exist. Some
clusters may have shallower mass profiles, because of substructure or
elliptical mass distributions, leading to our apparently
underestimating their L$_X$ for a given shear. However, it is not
straightforward to determine the effect such complications would 
have on the X-ray properties of the clusters and so we have chosen to
retain our simple description of the cluster potential.

As a baseline in comparing  models of the evolution of the mass and
X-ray emission in our distant clusters  we have  chosen to construct a
null-hypothesis.  We start with our analytical isothermal sphere model
of the cluster potential and calculate the shear produced as a function
of cluster mass, by lensing of a field population parameterized by the
same no evolution $N(z)$ used to correct the observed shear amplitudes
(\cite{irs91}).  The shear is calculated for a lens placed at $z= 0.4$
and integrated within the appropriate annulus.  We then determine the
equivalent X-ray temperature for our cluster mass, assuming the
intracluster gas is in hydrostatic equilibrium at the virial
temperature of the cluster potential.  We thus have  a relationship
between X-ray temperature and shear strength and using the locally
observed X-ray temperature--bolometric luminosity relation (T$_X
\propto $~L$_X^{0.30}$, \cite{ace91}; \cite{david93}) we can transform
this to an L$_X$--shear relation for the distant clusters.  To convert
from bolometric luminosity to L$_X$ in the 0.3--3.5 keV band we assume
a simple bremstrahlung spectrum for the cluster emission.   A power-law
fit to the final prediction using the form, $<\! g_1\! > = A_x
L_X^{\alpha_x}$ (with L$_X$ in units of h$^{-2} 10^{44}$ ergs sec$^{-1}$ in
the 0.3-3.5 keV band), gives $\alpha_x = 0.30\pm0.05$ and $A_x =
0.178^{+0.039}_{-0.032}$.  This prediction, based on the local
L$_X$--T$_X$ relationship, is termed our ``no evolution'' (NE) model.

A number of more elaborate evolutionary models have been discussed in
the literature (\cite{nk91}; \cite{eh}; \cite{wm}; \cite{rgb96}) to
explain the observed X-ray luminosity function (XLF) and its
evolution.  These models predict different forms for the cluster
temperature--luminosity relationship and we can compare these with our
null-hypothesis. The models are mostly based on analytical self-similar
models for cluster evolution using the Press-Schecter formalism.  We
will concentrate on the family of models discussed by Bower (1996), who
uses a simple parametrisation for the entropy evolution in the clusters
to predict the luminosity--temperature relationship for a number of
likely scenarios.  With an appropriate choice of the index of the mass
power spectrum, $n$,  these models can fit the local XLF and
L$_X$--T$_X$ relation, although the effect of $n$ on the evolution in the
L$_X$--shear relation is weak and so we have chosen to adopt $n=-1.5$.  

For a bolometric detector Bower predicts the following relation between
the characteristic temperature and bolometric luminosity of clusters:
T$_X \propto$  (1+z)$^{3\epsilon/11}$ L$_X^{4/11}$, where $\epsilon$
parameterises the entropy evolution.  We will restrict our discussion
to the the four values of $\epsilon$ discussed by Bower:  $\epsilon =
-3.7$, $-1.0$, $0.0$ and $1.0$.  Of these, $\epsilon = -3.7$
corresponds to the self-similar cluster evolution discussed by Kaiser
(1991), where the gas evolution is entirely driven by the cluster
growth.  $\epsilon=-1.0$ is an intermediate case, where the compression
of the gas during the cluster collapse is limited to some extent by the
minimum allowed entropy set by pre-heating.  $\epsilon = 0.0$ has a
constant entropy in the cluster core, and $\epsilon=1.0$ is a situation
where gas cooling in the cluster core dominates the evolution.  
We normalise these models by fitting each to the local $L_X/T_X$ data given
by Edge \& Stewart (1991) and, using the evolution of the characteristic
parameters, predict the $L_X$-shear correlations at $z$=0.4.

The relationship between the velocity dispersion of cluster galaxies
and the shear strength of the cluster is straightforward to determine,
assuming there is no velocity bias or similar effect in the cluster
members.   However, it should be noted that the velocity dispersions
typically sample a larger region of the cluster than the the X-ray
emission, which is centrally peaked and thus better matched to the
aperture of our lensing mass measurements.  For our sample definitions,
using the form, $<\! g_1\! > = A_v \sigma^{\alpha_v}$ (for $\sigma$ in
units of $10^3$ km sec$^{-1}$), gives $\alpha_v = 2.0$ and $A_v = 0.2$.
 
\section{Lensing Results}

In this section we now compare our cluster shear measurements with
other estimates of the cluster masses.  By determining the correlations
between these properties and comparing these to the equivalent,
locally-derived forms, we can investigate whether the physical mechanisms
underpinning the local relationships remain applicable in more distant 
clusters.  

\subsection{The Strong Lensing Regime}

We first undertook a visual search of the HST frames to uncover 
strongly-lensed features in the clusters.  Such features normally occur
when the mass density in the central regions of the cluster exceeds a
critical value. If our clusters have similar mass profiles we would thus
expect the presence of strongly-lensed features in the cluster cores to 
correlate with the shear strength of the clusters measured on larger scales. 
The results of this search are given in Table~1. 

We find candidate features in 65\% of our clusters, including both
giant arcs and multiply-imaged pairs. The presence of
strongly-lensed features was not a selection criterium for
the majority of our clusters (the exceptions being A2218, Cl0440$-$02 and
Cl0024+16) and thus their high rate of occurrence demonstrates the benefits
of working at high-resolution for the identification of lensed objects.
It is also interesting to note that the lensed features in 3 of the 7
strong-lensing clusters (A2218, AC118, Cl0054$-$27) indicate that these
clusters contain multiple mass components (e.g.\ Fig.~1).  These lensed
features and the more detailed view of the cluster mass distributions
they provide are dealt with in a separate article. We discuss below the
correlation between the presence of strongly-lensed features and the
mean cluster shear.

\subsection{The Weak Lensing Regime}

We now discuss our weak shear measurements in the context of 
L$_X$--$<\! g_1\! >$ and $\sigma$--$<\! g_1\!  >$ relationships.

\subsubsection{The L$_X$--$<\! g_1\! >$ Relation}

Fig.~2 shows the average tangential shear, within our adopted annular
region, versus the cluster's X-ray luminosity.  We detect a coherent
shear field, arising from gravitational lensing by the cluster
potential, in 11 of the 12 clusters in our study.  Those clusters which
show strongly-lensed features are indicated with filled symbols. Their
distribution in Fig.~2 shows that the average tangential shear
correlates well with the projected surface density in the inner $\sim
50$--100 h$^{-1}$ kpc. This would be expected if the clusters have
roughly similar mass profiles. The division between those clusters
capable of forming multiply-imaged features and those that cannot
provides a rough mass estimate for the central regions. For a
background source at $z\sim 1$, this region at $<\! g_1\! > \sim 0.1$,
is equivalent to a mass of $M(<50$h$^{-1}$kpc$) \sim 2 \times 
10^{13}$h$^{-1} M_\odot$.

A correlation is also apparent between the X-ray luminosity and the
mean shear, with the clusters with stronger shear fields showing higher
luminosities.  We fit a power-law, $<\! g_1\! > = A_x L_X^{\alpha_x}$,
to this correlation and determine best fit parameters for the slope,
$\alpha_x$, and normalisation, $A_x$. We obtain a best fit of $\alpha_x
= 0.58\pm0.23$ and $A_x = 0.074\pm0.017$.  However, the likelihood
contours for this fit (Fig.~3) shows that we can only realistically
constrain a combination of the slope and the normalisation.  The rms
scatter around this correlation is roughly a factor of $\sim 2$, which
is not much larger than would be expected from geometrical effects if
the clusters are a family of prolate ellipsoids with moderate axial
ratios.  A further indication that some of the observed scatter is
intrinsic comes from the cluster Cl0054$-$27. Fig.~2 shows this cluster
has a relatively low X-ray luminosity for its apparent shear
amplitude.  This is unlikely to be statistical scatter as the strong
shear is supported by the presence of strongly-lensed features in the
cluster center.  We suggest that this cluster is an example of a system
which is elongated along the line-of-sight, leading to a lower X-ray
luminosity than expected for its high mass surface density.  It should
be noted that standard X-ray deprojection, which assumes spherical
symmetry, applied to  such a system would lead to an incorrect estimate
of the cluster mass.

Fig.~2 shows the predicted upper bound relation expected in the case of
a 100\% shear measurement efficiency in our null-hypothesis (an
unevolving L$_X$--T$_X$ relationship). This line does approximately
define the upper bound for the sample. The estimated efficiency from
our simulations is 75\% (\S3) and the prediction in this case is also
shown. The data agrees with this prediction at the $\sim 99$\%
confidence limit.   Reducing the mean redshift of the background field
population used in our model would further improve the agreement.  For
example, a reduction in the median redshift of the field population by
10\% leads to a $\sim 10$\% increase in the corrected shear estimates
for the clusters, the effect is stronger for the more distant
clusters.  We conclude from Fig.~2 that there is no evidence for strong
evolution of the cluster L$_X$--temperature relationship between $z\sim
0$ and $z\sim 0.4$.

Next we examine the family of models from Bower (1996) discussed
earlier.  We plot the track followed by the models as a function of
$\epsilon$ on Fig.~3, marking the positions of the four cases listed
above.  Any values of $\epsilon \ls 0.5$ would provide as good a
description of the data as our NE model, although lower values are
preferred.  The 90\% confidence limits on the range of $\epsilon$ given
above are $-3.7 \leq \epsilon \leq -0.3$.  Very low values for
$\epsilon$, close to the self-similar case $\epsilon \sim -3.7$, are
excluded by the observed evolution in the cluster XLF (Kaiser 1991;
Bower 1996).  ~From the observed evolution in the XLF and temperature
function of distant clusters Bower (1996) derives limits of $-0.5 \leq
\epsilon \leq 3.8$ for an $n=-1.5$ power spectrum.  We can see that
both the lensing and X-ray data are roughly consistent with a value in
the range $\epsilon \sim -1$--0, if we include the possibility of
changing the index for the mass power spectrum.  This case represents
mild preheating of the X-ray gas, insufficient to produce a
constant entropy core at all epochs, but strong enough to significantly
alter the evolution. 

\subsubsection{The $\sigma$--$<\! g_1\!  >$ Relation}

When we study the relationship between $<\! g_1\! >$ and velocity
dispersion shown in Fig.~4 we come to a very different conclusion to
that obtained for the L$_X$--$<\! g_1\! >$ relation.  While the upper
bound predicted by the model does lie above the observations, the same
is also true when we include our best estimate of the observational
efficiency, 75\%, in contrast to Fig.~2 where this line passes through
the data points.  To achieve a similar agreement between the
observations and the model predictions as that seen for the
L$_X$--$<\!  g_1\! >$ relation we need to decrease the predicted shear
by a further 60\%, equivalent to an increase in the predicted velocity
dispersions of $\sim  40$--70\%.  

Given the relatively good agreement seen in the L$_X$--$<\!  g_1\! >$
correlation we suggest that this offset in the $\sigma$--$<\! g_1\!  >$
relation most likely arises from a general and systematic
over-estimation of our cluster velocity dispersions, by factors of
$\sim 40$\%.  This is the first time that a quantitative estimate of
the bias in velocity dispersion estimates for distant clusters has been
made.  However, we reiterate that the velocity dispersions are
typically measured over larger regions of the clusters than the other
mass estimates discussed here. We might expect that the quoted
discrepency would decrease if we restricted ourselves to galaxies
projected close to the cluster core, or for particular subsamples of
the clusters members (e.g.\ the E/S0 population).  The latter will be a
particular concern for the spectroscopic samples from Dressler \& Gunn
(1992), which were specifically aimed at identifying blue cluster
members, if the dynamics of such galaxies are not representative of the
cluster population as a whole.   Unfortunately, the velocity
information available at the present time is insufficient to allow us
to select the statistically reliable subsets needed to test our
suggestions on a cluster-by-cluster basis.  We can, however, combine
the velocity information for a number of clusters and determine how the
typical dispersion would change if we restricted our sample to only the
``passive'' galaxies. Using the redshift and spectral information from
Dressler \& Gunn (1992) we find that the typical cluster dispersion for
the passive population is $\sim 25$\% less than the whole sample.  Such
a decrease would obviously go a long way to reconciling the observed
dispersion with the lensing predictions.  A related possible cause of
the systematic overestimation of the velocity dispersions arises from
the difficulty of identifying outliers and substructures in the
distributions for individual clusters when dealing with samples of only
$\sim 30$ members per cluster (\cite{carl96}).  An indication of the
change in the velocity dispersion when using a more complex iterative
rejection of interlopers and no color selection for the galaxies, comes
from a comparison of our estimated dispersion for Cl0016+16 with that
from the slightly larger sample amassed for this cluster by the CNOC
collaboration (\cite{carl96}).  The CNOC group estimate $\sigma=1243
$km sec$^{-1}$, a value 40\% lower than the value given in Table~1,  in line
with our estimate of the offset given above.   

The extent of the overestimation of the cluster dispersions in our
sample, which all have apparently well sampled distributions, and its
variation with the sample definition may be also indicating that
{\it most} distant clusters are dynamically unrelaxed, containing
coalescing and unvirialised mass sub-components as well as more general
in-fall (\cite{hcf96}).  Direct evidence of such sub-components comes from the
strongly-lensed features in the most massive clusters. In $\sim 40$\%
of these clusters, the morphologies of the strongly-lensed features
indicate that the mass distributions  contain multiple components on
projected scales of 100--$300 $h$^{-1}$ kpc. This leads us to suggest
that these massive distant clusters are observed in a period of rapid
growth.  

\subsection{Mass to Light Ratios}

We next wish to convert the shear estimates for our clusters into
masses and combine these with measurements of the luminosity of the
cluster population within a similar aperture to determine the mass to
light ratios for the various clusters.  Using a singular isothermal
sphere model for the clusters and the no evolution $N(z)$, we estimate
that an observed shear of $<\!  g_1\! > = 0.1$ measured inside a 200
h$^{-1}$ kpc aperture in a cluster at a redshift of $z=0.4$ corresponds
to a projected mass within the same radius of $M = 9.8 \times 10^{13}
$h$^{-1} M_\odot$.  We use this value along with the observed shears to
give the cluster aperture masses listed in Table~2.  

Weak lensing observations have been published for the cluster Cl0016+16
from ground-based observations (\cite{irs95a}).  Converting their
values from a $600 $h$^{-1}$ kpc diameter aperture to that used here we
find that their lower limit to the cluster mass inside a radius of $200
h^{-1}$ kpc is $M\geq 0.9 \times 10^{14} $h$^{-1} M_\odot$ and their
best estimate for the mass is $M\sim 2.9 \times 10^{14} $h$^{-1}
M_\odot$.  This is in reasonable agreement with our value of $M\sim
(1.87\pm0.64) \times 10^{14} $h$^{-1} M_\odot$ from Table~2,
particularly given the difficulties of the ground-based observations
and the large corrections which have to be applied to them.

To estimate the luminosity associated with the cluster galaxy
population we must first attempt to separate and correct for the
contamination by field galaxies.  We have chosen to do this using the
morphologically classified field counts from the HST Medium Deep Survey
(MDS, Griffiths et al.\ 1994).  By similarly classifying the galaxy
populations in our cluster fields we can then correct the individual
galaxy classes for field contamination to determine total luminosities
for the various cluster populations.  The morphological catalogs of
galaxies in these clusters are given in Smail et al.\ (1996) and Couch
et al.\ (1996) to a limit of $I_{814}=23.0$ or $R_{702}=23.5$.   These
samples have been visually classified onto the revised Hubble scheme,
similar to that used by the MDS, in a manner described in the
associated papers.  We fit power laws to the differential number counts
for the different morphological classes (E, S0, Sab, Scd, Irr) from the
MDS catalogs and use these to subtract off the counts in the cluster
fields.  Having done this we then convert the magnitudes of the
resulting cluster populations to $M_V$ using the relevant K correction
for the spectral energy distributions (SED) of the different
morphological classes (assuming that the SED of a particular
morphological type is not a function of epoch).   These
field-subtracted luminosity functions are binned into two samples:  E
and all galaxies (E-Irr), which are then integrated down to a fixed
absolute magnitude ($M_V=-18.5$) and corrected for both the proportion
of the population missed due to the annulus falling outside the frame
(assuming all the galaxy populations follow an isothermal distribution)
and the fraction of light missed from galaxies below the adopted
magnitude limit (for this we use the extrapolate the fits to the
cluster elliptical and spiral LFs given in Smail et al.\ 1996) to
obtain total luminosities.   

We have chosen to calculate total luminosities for the elliptical
population as well as for the whole cluster population.  There are
three reasons for this: (1) elliptical galaxies are a large fraction of
the cluster population, but only a relatively small proportion of the
field population and thus the field contamination of this luminosity
estimate ought to be lower than a sample comprising both ellipticals
and spirals; (2) the majority of the clusters included in this study
contain large populations of blue star-forming galaxies which are
absent from similar environments today consequently the blue luminosity
of the whole cluster population may be artificially raised; (3) the
evolution of the elliptical cluster population is better understood
than the clusters spirals (\cite{rse96}; \cite{ajb96}), this allows us
to robustly predict the luminosity of the elliptical population at the
current epoch.

We show the relationship between cluster aperture mass and luminosity
for the various sample definitions in Fig.~5 and list the values in
Table~2.  For the whole cluster population, we
obtain a median mass to light ratio of $M/L^{\rm all}_V =
180^{+140}_{-80} $h$ (M/L_V)_\odot$, where the limits are the quartile
points of the distribution.   That for the ellipticals only is 
$M/L^{\rm E}_V = 330^{+210}_{-110} $h$
(M/L_V)_\odot$.  Correcting for the observed evolution in the
elliptical galaxy population to the individual clusters ($\delta M_V
\sim -0.6$ to $z=0.4$ c.f.\ \cite{ajb96}) we obtain M/L$^{\rm E}_V (z=0) =
620^{+250}_{-240} $h$ (M/L_V)_\odot$.  For comparison M/L$_V =
1400$--$1600  $h$ (M/L_V)_\odot$ is required locally for closure density
(\cite{bt87}).

Clearly the most interesting of the three  mass to light ratios listed
above is that for the whole cluster population.  To convert this to a
present-day value we must first determine the luminosity evolution of
the whole cluster population from $z=0.4$ to the present day.
Unfortunately, studies of the Butcher-Oemler effect in distant clusters
(\cite{cess}; \cite{dobg}) show that the evolutionary history of the
spiral galaxies, which comprise 25--65\% of the total cluster
luminosity in these systems, is likely to be complex and so we do not
pursue this approach. Instead we turn to the elliptical population,
whose evolution is better understood.  Here, however, we are only
dealing with a proportion of the total cluster luminosity and so we
will only be able to derive an upper limit for the cluster M/L at the
present epoch.  Including the fading of this population between the
observed epoch and the present day and assuming that this mass to light
ratio is representative of the global value we have  $\Omega_0 \leq
0.4\pm0.2$.  

Looking at the scatter in the various measurements we note that the
luminosity of the elliptical population is better correlated  with the
cluster aperture mass than the luminosity of the whole cluster
population.  Moreover, the relative scatter for the M/L$^{\rm E}$ is only
$\sim 40$\%, very similar to the median error in the cluster masses
from the lensing analysis.  This similarity leaves open the possibility
that in the majority of the clusters all the observed scatter in the
M/L$^{\rm E}$ arises from measurement errors and hence that the cluster
M/L$^{\rm E}$ ratio is a constant.  Such a conclusion would obviously have
profound implications for the formation mechanism of cluster
ellipticals.  However, we also note that some clusters in Fig.~5
populate a tail to considerably higher M/L's.  These clusters tend to
be more massive than the mean in our sample, and some variation in
cluster property (e.g.\ mass profile) with mass may therefore be
driving their anomalous M/L's.  

Finally, we give an illustration of the unique insights which
gravitational lensing provides into the structure and evolution of
clusters.  In Fig.~6 we plot the spiral fractions of the clusters from
Table~2 ($f_{sp}$) against their aperture masses.  This figure exhibits
a transition between the spiral fractions of low and high mass
clusters, the change over apparently occuring at a mass of $M\sim
2 \times 10^{14} $h$^{-1} M_\odot$.  Above this mass the central regions of
the clusters ($\leq 0.5 $h$^{-1}$ Mpc) are relatively devoid of spiral
galaxies.   This figure bears a striking resemblance to Fig.~16 of
Allington-Smith et al.\ (1993), where they plot blue fraction, $f_B$,
versus richness for local groups and clusters.  Allington-Smith et al.\
found a sharp decrease in the blue-fraction for the richest systems
(with total luminosities above $L_V \sim 3 \times 10^{12} $h$^{-2} L_\odot$,
Oemler (1992)).  However, when they constructed a similar plot for
distant groups and clusters (their Fig.~19) they saw little evidence
for a similar decrease in $f_B$ with increasing richness.  The
existence of a sharp discontinuity in our morphology-environment
relation (when determined from $f_{sp}$ and cluster mass) maybe
indicating their use of total richness as an indication of system mass
undermined their analysis.  Clearly more data is required to
investigate the feature in the morphology-environment relation, and to
untangle it from the strong redshift evolution observed in the relation
(Fig.~6, \cite{cess}; \cite{dobg}).  Nevertheless, we stress that the
impact of adding lensing masses to the list of observable
characteristics of distant clusters may be felt across a wide range of
research fields.

\section{Conclusions}

Our study represents the first analysis of the weak gravitational
lensing signal in a large and diverse sample of distant clusters. The major
advantage of our study has been the use of HST to detect the weak
shear, where we gain over ground-based telescopes by the high
efficiency for recovering the true shear. 

Summarising the main conclusions of our survey:

\noindent{$\bullet$} We have measured a tangential alignment of the
images of faint background galaxies in 11 out of a sample of 12 distant
clusters imaged with HST. The high detection rate demonstrates the
power of high-resolution imaging in this type of study.    

\noindent{$\bullet$} We show that the presence of strongly-lensed
features within the cores of a large fraction of our clusters
correlates well with the shear signal measured on larger scales from
the more weakly distorted arclets. 
  
\noindent{$\bullet$} We find a reasonable correlation between the X-ray
luminosities of our clusters  and their masses, as estimated from the
gravitational shear fields in the central $\sim 0.5$ h$^{-1}$ Mpc of
each cluster.  The scatter about this relation is only a factor of
two -- not much larger than that expected from projection effects  in a
family of randomly oriented ellipsoidal clusters.  

\noindent{$\bullet$} We compare the L$_X$--shear data with a prediction
made on the basis of the local L$_X$--T$_X$ relation and isothermal 
mass and gas distributions in the clusters. We find that the local 
relation is an adequate description of the distant cluster sample, after 
allowance has been made for a number of observational effects and 
uncertainties. We thus conclude that there is no strong evidence for
evolution in the L$_X$--$T_X$ relationship since $z\sim 0.4$.

\noindent{$\bullet$} The observed evolution in the L$_X$--T$_X$
relation can be reproduced by models introducing a modest initial
entropy into the gas prior to cluster formation, possibly resulting
from pre-heating by AGN or galactic winds.   Such a family of models
has also been discussed in regard to the evolution of the XLF
(\cite{fjc95}; Bower 1996). 

\noindent{$\bullet$} The velocity dispersions for those clusters with
adequate data show that the typical cluster has a higher measured
dispersion than expected on the basis of both their weak lensing and
X-ray luminosities.  We suggest that this arises from an inability to
identify outliers and unvirialised substructures within the clusters in
the relatively modest spectroscopic samples.  By restricting our sample
to only the passive, red populations in the clusters we can reduce
these problems, and we illustrate that dispersions are reduced by a
sizeable factor.  Nevertheless, the strength and ubiquity of these
biases raises doubts over the usefulness of virial analysis of distant
clusters, when restricted to only modest samples of members.  

\noindent{$\bullet$} We convert our shear measurements into estimates
of the mass in the central regions of the clusters.  We then combine
these with measurements of the luminosities of various samples of
cluster galaxies to determine mass to light ratios for the cluster.  By
concentrating on the well understood elliptical population we are able
to derive a limit of M/L$_V \leq 620^{+250}_{-240} $h$ (M/L_V)_\odot$ 
for the clusters at $z=0$, equivalent to $\Omega_0 \leq 0.4$.  

\noindent{$\bullet$} We illustrate the possible impact of observations
such as these on studies of distant clusters by showing the correlation
between the spiral fractions and the cluster masses.  This shows that
the morphological mix in the clusters undergoes a rapid change for
clusters with masses above $M \sim 2 \times  10^{14} $h$^{-1}
M_\odot$. 

Studies such as the one presented here will be enlarged in the future
through high-resolution imaging of larger samples. These should include
a better-defined subset of massive distant clusters to investigate the
intrinsic scatter in their X-ray/lensing properties arising from
geometrical effects, as well as quantify the rate of occurrence of
substructure and hence their rate of growth. An extended study would
also benefit from the inclusion of more low mass systems to constrain
the slope of the L$_X$--mass relationship at moderate redshift, as well
as the morphology-mass relationship which underlies the Butcher-Oemler
effect.  Extending the analyzes to higher redshift will provide new
constraints on the form of any evolution in the L$_X$--mass relation.
This paper has demonstrated that weak lensing with HST provides a new
and highly promising means for measuring the evolution of large scale
systems in the universe.

\section*{Acknowledgements}
We wish to thank Ray Lucas at STScI for his enthusiastic help which
enabled the efficient gathering of these HST observations.  We
especially thank the anonymous referee for their constructive comments
and suggestions which helped to significantly improve and clarify this
paper.  We also thank Richard Bower, Alastair Edge, Vincent Eke,
Jean-Paul Kneib, Jordi Miralda-Escud\'e and John Mulchaey for many
useful conversations and assistance. IRS acknowledges support from a
Carnegie Fellowship and RSE and RMS acknowledge support from the
Particle Physics and Astronomy Research Council.  WJC acknowledges
support from the Australian Department of Industry, Science and
Technology, the Australian Research Council and Sun Microsystems.

\smallskip
\vfil\eject

\section*{Tables}

\noindent{\bf Table 1} The cluster sample used in our analysis,
including field identification, cluster redshift, filter passband,
total exposure time, X-ray luminosity in the 0.3--3.5 KeV band,
velocity dispersion (where known), angular scale, surface brightness
limit (the 1$\sigma$ fluctuation in the sky flux in a 1 sq.\ arcsec
aperture) and whether candidate strongly-lensed features are present.

\noindent{\bf Table 2} Properties of the cluster galaxy populations.
These are quoted as values corrected to an aperture of $400 h^{-1}$ kpc
diameter.  We give the estimated cluster mass from the lensing
analysis, the integrated luminosity of the elliptical population,
integrated luminosity of the whole cluster population, integrated
luminosity of the elliptical population evolved to the present day and
the equivalent mass to light ratios.   These are all corrected for
luminosity in galaxies lying below the magnitude limits of the
morphological samples and for those regions of the aperture falling off
the frame.  The final column gives the spiral fractions in the
clusters, as a proportion of the whole population brighter than
$M_V=-18.5$.  We omit those clusters without morphologically typed
samples or which are undetected in our shear measurements.  The
luminosities have not been corrected for reddening.

\vfil\eject

\hbox{~}\vskip0.5truein
\centerline{\hbox{
\psfig{figure=lxmass_fig1a.ps}
}}

{\bf Figure 1.} The shear fields, indicated by the vectors, for four of
the stronger lenses in the sample. The clusters illustrated span a wide
range in redshift and the shear fields show strong coherence around the
lens center, which always coincides with the brightest member (marked
by +). The vectors represent independent data points, each containing
roughly $\sim 10$ faint galaxies. A perturbation in the shear field
associated with a bright D galaxy just off the top-right of the frame
is visible in Fig.~1(b). The scale is arcseconds and the vector at the
top-left of each frame represent a 20\% shear.  (a) A2218 $z=0.17$,
(b) AC118 $z=0.31$, (c) 3C295 $z=0.46$ and (d) Cl0016+16 $z=0.55$.

\vfil\eject

\hbox{~}\vskip0.5truein
\centerline{\hbox{
\psfig{figure=lxmass_fig1b.ps}
}}
\centerline{\bf Figure~1b}

\vfil\eject

\hbox{~}\vskip0.5truein
\centerline{\hbox{
\psfig{figure=lxmass_fig1c.ps}
}}
\centerline{\bf Figure~1c}

\vfil\eject

\hbox{~}\vskip0.5truein
\centerline{\hbox{
\psfig{figure=lxmass_fig1d.ps}
}}
\centerline{\bf Figure~1d}

\vfil\eject

\hbox{~}\vskip0.5truein
\centerline{\hbox{
\psfig{figure=lxmass_fig2.ps}
}}

{\bf Figure 2.} The correlation between the cluster X-ray luminosity
and the mean shear strength, $<\! g_1\! >$. The error bars are $1\sigma$ 
boot-strap estimates and the solid line shows the best fit relationship for 
the data. The dotted line indicates the upper limit expected, assuming 100\% 
measurement efficiency, in the case of our simple ``no evolution'' model. 
The dashed line represents a 75\% efficiency.  Filled symbols denotes those 
clusters which have candidate strongly-lensed features.

\vfil\eject

\hbox{~}\vskip0.5truein
\centerline{\hbox{
\psfig{figure=lxmass_fig3.ps}
}}
\vskip-0.3truein

{\bf Figure 3.} The best fitting parameters for a power-law description
of the L$_X$--$<\! g_1\! >$ correlation, the contours are 50\%, 90\%,
95\%, 99\% and 99.9\% confidence limits.  The vertical bar shows the
limit on the normalisation from the ``no evolution'' model, determined
by assuming 100\% measurement efficiency in our shear estimation.  All
observational effects will tend to drive down the predicted shear
amplitude, pushing the prediction further to the left of this line.
The point with error bars indicates the NE prediction assuming our best
estimate of 75\% measurement efficiency. The predictions of the other
models discussed in the text as a function of the parameter $\epsilon$
lie along the track plotted.  The points for $\epsilon= -3.7$, $-1.0$,
$0.0$ and $1.0$ are marked, these all include a 75\% signal
degradation.

\vfil\eject

\centerline{\hbox{
\psfig{figure=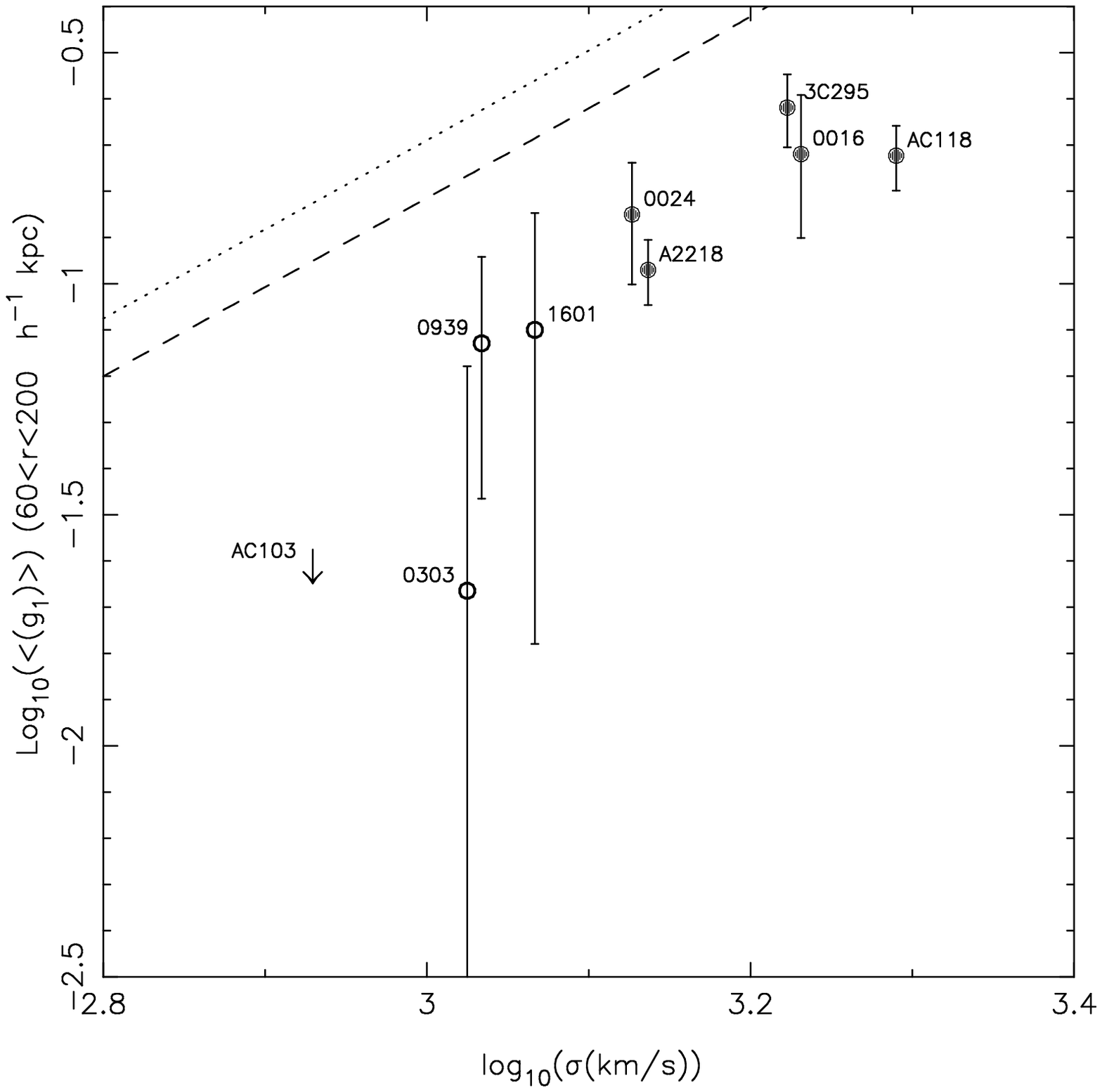}
}}

{\bf Figure 4.} Correlation between cluster velocity dispersion and
$<\! g_1\! >$ for those clusters with redshifts for at $> 20$ members.
Error bars are $1\sigma$. The dotted line shows the theoretical
prediction assuming 100\% measurement efficiency; the dashed line
represents an efficiency of 75\% as expected from simulations.  Filled
symbols denote those clusters with candidate strongly-lensed features.
The arrow marks the $1\sigma$ upper limit on the AC103 shear
measurement.

\vfil\eject

\hbox{~}\vskip0.5truein
\centerline{\hbox{
\psfig{figure=lxmass_fig5a.ps}
}}

{\bf Figure 5.} Plots of the total cluster mass versus integrated light
in various cluster galaxy populations (both corrected to a $400 h^{-1}$ kpc
diameter aperture.  These are shown for: (a) the whole cluster
population, (b) just those galaxies classed as elliptical on the basis
of their morphology, (c) the elliptical population evolved to the
present day.  All of the cluster luminosities have been corrected for
field contamination, losses from regions of the aperture falling
outside the frame and galaxies missed owing to the adopted magnitude
limits as is described in the text.  The solid lines show the median
mass to light ratio of each sample.  The histograms in each panel show
the distribution of the mass to light ratios.

\vfil\eject
\hbox{~}\vskip0.7truein
\centerline{\hbox{
\psfig{figure=lxmass_fig5b.ps}
}}

\centerline{\bf Figure~5b}

\vfil\eject
\hbox{~}\vskip0.7truein
\centerline{\hbox{
\psfig{figure=lxmass_fig5c.ps}
}}

\centerline{\bf Figure~5c}

\vfil\eject

\hbox{~}\vskip0.5truein
\centerline{\hbox{
\psfig{figure=lxmass_fig6.ps}
}}

{\bf Figure 6.}  The number fraction of the cluster population
($f_{sp}$) brighter than $M_V=-18.5$ which have spiral morphology
(Sab-Irr), plotted against the lensing estimate of the cluster aperture
mass.   There appears to be a sharp transition in the spiral fraction
of the clusters at masses around $M \sim 2 \times 10^{14} h^{-1} M_\odot$.
However, we caution that owing to the strong redshift evolution known
to exist in $f_{sp}$ (c.f.\ \cite{cess};\cite{dobg}) this figure is not
simple to interpret.  The position of the $z=0.17$ cluster A2218 in
this figure probably reflects  redshift evolution,  rather than a real
difference.

\vfil\eject

{\singlespace
\begin{center}
\begin{tabular}{lccccccccc}
\multispan{9}{\bf \hfil Table 1 \hfil }\\
\noalign{\smallskip}
\hline
\noalign{\smallskip}
{Cluster} & {\it z} & & {T$_{\rm exp}$} & & {L$_{\rm X}$} &
$\sigma$ \hfil &  {kpc/$''$} & $\mu (1\sigma)$ & {Strong} \cr
& & {\scriptsize F555W} & {\scriptsize F702W} & {\scriptsize F814W} & {\scriptsize (0.3--3.5) h$^{-2}$ 10$^{44}$} & {\scriptsize km sec$^{-1}$ [N]}& {\scriptsize h$^{-1}$} & {\scriptsize  mag/arcsec$^{2}$} & {Lensing} \cr
\noalign{\smallskip}
\noalign{\hrule}
\noalign{\smallskip}
A2218       & 0.17 & ---  &  6.5 & ---  & 2.85 & 1370 [50]&  1.88 & 27.2 & yes \cr
Cl0440$-$02 & 0.19 & ---  & 18.4 & ---  & 1.00 & ---\hfil &  2.03 & 27.1 & yes \cr
AC118       & 0.31 & ---  &  6.5 & ---  & 3.90 & 1950 [34]&  2.80 & 27.2 & yes \cr
AC103       & 0.31 & ---  &  6.5 & ---  & ---  &  850 [29]&  2.80 & 27.2 & no  \cr
Cl0024+16   & 0.39 & ---  &  --- & 13.2 & 0.55 & 1339 [33]&  3.17 & 27.5 & yes \cr
Cl0939+47   & 0.41 & ---  & 21.0 & ---  & 1.05 & 1081 [31]&  3.26 & 27.3 & yes \cr
Cl0303+17   & 0.42 & ---  & 12.6 & ---  & 0.55 & 1079 [21]&  3.29 & 27.9 & no  \cr
3C295       & 0.46 & ---  & 12.6 & ---  & 3.20 & 1670 [21]&  3.43 & 27.1 & yes \cr
Cl0412$-$65 & 0.51 & 12.6 & ---  & 14.7 & 0.08 & ---\hfil &  3.59 & 27.5 & no  \cr
Cl1601+43   & 0.54 & ---  & 16.8 & ---  & 0.35 & 1166 [27]&  3.67 & 28.3 & no  \cr
Cl0016+16   & 0.55 & 12.6 & ---  & 16.8 & 5.88 & 1703 [30]&  3.67 & 26.7 & yes \cr
Cl0054$-$27 & 0.56 & 12.6 & ---  & 16.8 & 0.25 & ---\hfil &  3.76 & 27.4 & yes \cr
\end{tabular}
\bigskip

\begin{tabular}{lrcrcrrrrr}
\multispan{9}{\bf \hfil Table 2 \hfil }\\
\noalign{\smallskip}
\hline
\noalign{\smallskip}
{Cluster} & {M \hfil} & {L$^{\rm E}_V$} & {L$^{\rm all}_V$\hspace{5truemm}} &
{L$^{\rm E}_V (z=0)$} &   
{M/L$^{\rm E}_V$} & {M/L$^{\rm all}_V$}
& {M/L$^{\rm E}_V (z=0)$} & {f$_{\rm sp}$} \cr
& {\scriptsize 10$^{14}$ h$^{-1}$ M$_\odot$} & {\scriptsize 10$^{11}$ h$^{-2}$ L$_\odot$} & {\scriptsize 10$^{11}$ h$^{-2}$ L$_\odot$} &
{\scriptsize 10$^{11}$ h$^{-2}$ L$_\odot$} & 
{\scriptsize $h$ (M/L)$_\odot$} & {\scriptsize $h$ (M/L)$_\odot$} & {\scriptsize $h$ (M/L)$_\odot$} & \cr
\noalign{\smallskip}
\noalign{\hrule}
\noalign{\smallskip}
A2218       &  1.05$\pm$0.19 & 2.57 &  3.39\hspace{5truemm} & 2.04 &  410 & 310 &  520\hspace{5truemm} &  0.17  \cr
AC118       &  1.85$\pm$0.32 & 2.72 &  5.06\hspace{5truemm} & 1.77 &  680 & 370 & 1040\hspace{5truemm} & 0.21  \cr
Cl0024+16   &  1.38$\pm$0.37 & 5.15 &  9.31\hspace{5truemm} & 3.00 &  270 & 150 &  460\hspace{5truemm} & 0.40 \cr
Cl0939+47   &  0.73$\pm$0.41 & 2.15 &  6.11\hspace{5truemm} & 1.22 &  340 & 120 &  600\hspace{5truemm} & 0.46 \cr
Cl0303+17   &  0.22$\pm$0.45 & 1.25 &  2.54\hspace{5truemm} & 0.70 &  170 &  80 &  310\hspace{5truemm} & 0.46 \cr
3C295       &  2.35$\pm$0.38 & 2.61 &  7.16\hspace{5truemm} & 1.38 &  900 & 330 & 1700\hspace{5truemm} & 0.28 \cr
Cl0412$-$65 &  0.25$\pm$0.41 & 1.75 &  3.72\hspace{5truemm} & 0.87 &  150 &  70 &  290\hspace{5truemm} & 0.47 \cr
Cl1601+43   &  0.77$\pm$0.66 & 2.50 &  4.05\hspace{5truemm} & 1.19 &  440 & 190 &  650\hspace{5truemm} & 0.46 \cr
Cl0016+16   &  1.87$\pm$0.64 & 5.68 & 10.62\hspace{5truemm} & 2.66 &  330 & 180 &  700\hspace{5truemm} & 0.21 \cr
Cl0054$-$27 &  1.71$\pm$0.64 & 1.50 &  4.31\hspace{5truemm} & 0.69 & 1140 & 400 & 2480\hspace{5truemm} & 0.42 \cr

\end{tabular}

\end{center}
}

\vspace*{.5cm}

\end{document}